\documentclass[%
 reprint,
superscriptaddress,
 amsmath,amssymb,
 aps,
]{revtex4-2}

\usepackage{graphicx}% Include figure files
\usepackage{dcolumn}% Align table columns on decimal point
\usepackage{bm}% bold math
\usepackage{setspace}%\usepackage{dcolumn}
\usepackage{bm}
\usepackage{caption}
\usepackage{subcaption}
\usepackage[linguistics]{forest}
\usepackage{booktabs}
\usepackage{color, colortbl}
\usepackage{amsmath}
\usepackage{physics}
\usepackage{mhchem}[version=4]
\usepackage{setspace}
\usepackage{soul}
\usepackage{color,soul}
\usepackage{threeparttable}
\usepackage{import}

\begin{document}

\title{First principles computations of the Stark shift of a defect-bound exciton: the case of the T center in silicon}

\author{Louis Alaerts}
\affiliation{Thayer School of Engineering, Dartmouth College, Hanover, NH 03755, USA}
\author{Yihuang Xiong}
\affiliation{Thayer School of Engineering, Dartmouth College, Hanover, NH 03755, USA}
\author{Sin\'ead M. Griffin}
\affiliation{Molecular Foundry, Lawrence Berkeley National Laboratory, Berkeley, CA 94720, USA}
\affiliation{Materials Sciences Division, Lawrence Berkeley National Laboratory, Berkeley, CA 94720, USA}
\author{Geoffroy Hautier}
\affiliation{Thayer School of Engineering, Dartmouth College, Hanover, NH 03755, USA}
\affiliation{Department of Materials Science and NanoEngineering, Rice University, Houston, TX 77005,USA}

\date{\today}% It is always \today, today,
             %  but any date may be explicitly specified

\begin{abstract}

The T center in silicon has recently drawn a lot of attention for its potential in quantum information science. The sensitivity of the zero-phonon line (ZPL) to electrical field was recently investigated by a combination of different experimental methods but there is still no first principles study on the Stark shift of the T center. Dealing with the defect-bound exciton nature of the excited state is particularly challenging using density functional theory because of the large spatial delocalization associated with the wavefunction. Here, we tackle this issue by performing a convergence study over the supercell size. We obtain an exciton binding energy of 28.5meV, in good agreement with experimental results. We then calculate the Stark shift through the dipole moment change of the ZPL transition of the T center using the modern theory of polarization formalism and find a modest linear coefficient of $\Delta \mu$=0.79D along X and $\Delta \mu$=0.03D along Y. We discuss our results in light of the recent experimental measurements of the Stark shift. Our analysis suggests that bound-exciton defects could be particularly sensitive to local field effect as a result of their large spatial extent. 

\end{abstract}

%\keywords{Suggested keywords}%Use showkeys class option if keyword
                              %display desired
\maketitle

%\tableofcontents

\section{Introduction}
Color centers in semiconductors have emerged as promising spin-photon interfaces for quantum information science (QIS) applications. Such quantum defects have already proved key for the development of quantum networks for their use as single-photon generation or remote entanglement of electron spin qubits \cite{Hensen2015, Knaut2024}. Among the various hosts for quantum defects, silicon is of particular interest owing to its high technological maturity, facilitating integration into electronic and nanophotonic devices \cite{Yan2021, Komsa2014, Deabreu2023, Day2024, Day2025, Dobinson2025}. The most attractive spin-photon interface defect in silicon is currently the T center which consists of two carbons with one bonded to an hydrogen, substituting a single silicon site, with point symmetry C$_{1h}$ (see Figure \ref{fig:T_center_basis} (A)). The T center possess sought-after features, such as a doublet ground state, emission in the telecom O-band and long spin coherence times \cite{Bergeron2020, Higginbottom2022, Dhaliah2022}. The sensitivity of the zero-phonon line (ZPL) emission to electric field or Stark shift is a very important property for practical quantum defects. Large sensitivity to electric field is detrimental for quantum defects because it often translates to large spectral diffusion, thus limiting the optical coherence of the emitter \cite{Wolfoicz2021}. The effect of electric field on the T center emission has begun to be explored experimentally both directly or indirectly through recent integration of T center's in p-i-n diodes \cite{Clear2024, Dobinson2025, Day2025}. While the Stark shift has been computed from first principles on the NV center in diamond \cite{Udvarhelyi2019, Lopez2024, Alaerts2024}, there is no full first principles evaluation of the Stark shift of the T center within a supercell approach. 

Here, we evaluate from first principles the Stark shift of the ZPL using the modern theory of polarization. We carefully converge all quantities with supercell size as the bound exciton nature of the excited state in the T center leads to extended wavefunctions. We compare our results to experimental data and discuss the nature of the Stark shift in bound exciton defects. Beyond the T center, this paper points out to methodological challenges and physical mechanisms proper to defect-bound excitons in general.

\section{\label{sec:level1} Results}
\subsection{Defect-bound exciton binding energy}
\begin{figure}[h]
    \centering
    \includegraphics[width=1\linewidth]{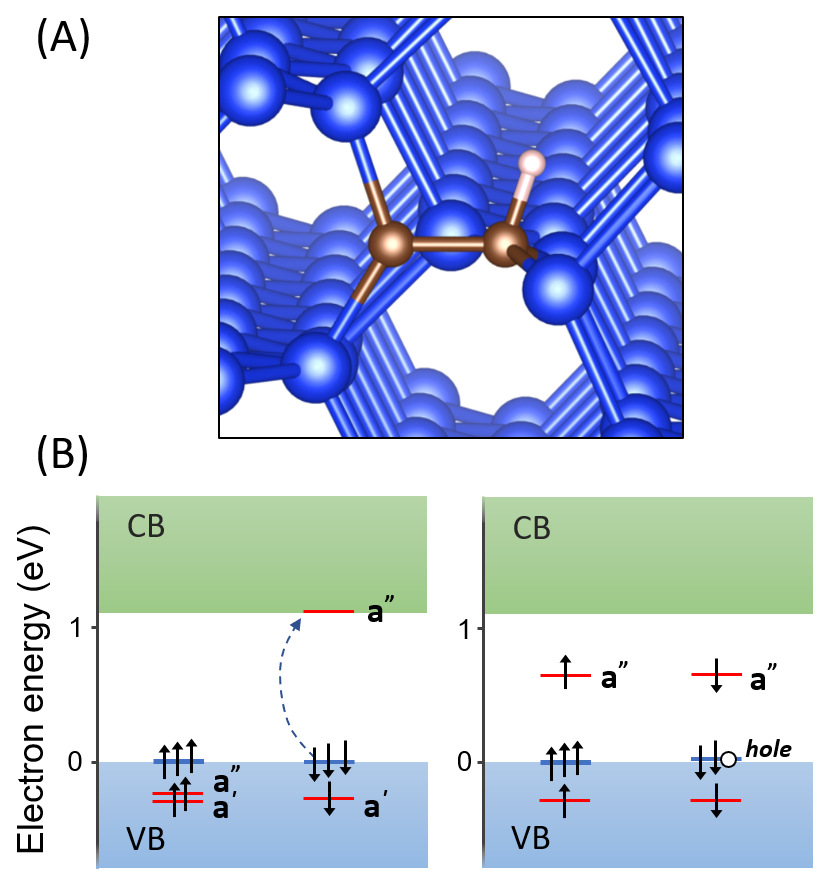}
    \caption{(A) Molecular model of the T center (B) (left) Kohn-Sham eigenvalues of the ground state of the T center obtained from a HSE calculation on a 6x6x6 supercell and, (right) same for the excited state. The Kohn-Sham eigenvalues are in agreement with our previous work \cite{Dhaliah2022}.}
    \label{fig:T_center_basis}
\end{figure}

The electronic structure and molecular model of the T center are shown in Figure \ref{fig:T_center_basis}. The excited state of the T center is a defect-bound exciton resulting from the excitation of an electron from the valence band to the a'' unoccupied state \cite{Dhaliah2022}. The delocalized hole in the valence band is attracted by the negatively charged T center.  This is a \textit{pseudo-acceptor defect-bound exciton} by analogy with group III impurities such as boron which are acceptors and introduce shallow defects that are occupied by an electron while a hole is present in a valence band-like state. We note that we only consider here the first excited state called T$_0$.

The first quantity we compute is the exciton binding energy for the excited state. Defined as the energy needed to dissociate the hole from the charged defect, it has been first measured by temperature-dependent PLE spectroscopy in the range of 22.5eV to 32meV \cite{Irion1985, Safonov1996} for the T center. A more recent estimation obtained by matching the bound-exciton transition levels to the ones of boron gave a similar value of 35meV \cite{Bergeron2020}. For comparison, the acceptor state of group III impurities B, Al, Ga and In have a binding energy (the energy needed to transfer the hole into the valence band) of 45.71, 70.18, 74.05 and 156.90meV respectively \cite{Pajot2010}. Using density functional theory (DFT), the exciton binding energy can be computed using the ZPL energy and the neutral to negative charge transition energy. The ZPL corresponds to the energy of the bound state while the charge transition energy is the energy of an entirely delocalized hole and of a negatively charged T center. We recently used this approach on the T center and obtained a value of 85meV \cite{Dhaliah2022}, clearly above the 22 to 35meV measured experimentally. This overestimation is caused by the delocalized nature of the exciton, which requires a supercell large enough to fully encompass the spatial extent of the hole wavefunction, yet also avoiding spurious exciton-exciton interaction due to periodicity. This effect has been previously observed in the study of the binding energy of group III impurities for which a cell size of up to 64,000 atoms was needed \cite{Wang2009, Zhang2013}, far larger than the standard supercell of 512 atoms used in our previous work \cite{Dhaliah2022}. Recently, Swift et al. calculated the binding energies of shallow acceptors in silicon by first principles, performing a convergence study focused on the supercell size \cite{Swift2020}. Here, motivated by the high degree of similarity between the excited state of the T center and group III acceptors/donors, we follow a similar route. In our previous work, we used the charge transition and ZPL energy to obtain the binding energy but this approach requires three calculations: the ground state, excited state and the negatively charged T center calculation. On the other hand, Swift et al. showed that binding energies calculated using the Kohn-Sham eigenvalues \cite{Swift2020} are accurate, alleviating the need for the negatively charged calculation. Within this approach, the binding energy is given by:

\begin{equation}
    E_B = \varepsilon_{KS-hole} -  \varepsilon_{VBM} + e \Delta V
\end{equation}

Where $\varepsilon_{KS-hole}$ is the position of the empty state in the valence band, $\varepsilon_{VBM}$ is the valence band maximum and e$\Delta V$ is a correction term that ensures that the position of the VBM are aligned in the bulk and defect calculations. Because the T center is a pseudo-acceptor, $\varepsilon_{KS-hole}$ is evaluated by a $\Delta$-SCF calculation in which the T center level $a''$ has been constrained to be occupied. 
The position of the VBM in a defect calculation will be shifted because of finite size effects and $\varepsilon_{VBM}$ should therefore be evaluated by a separate calculation on a pristine supercell of the same size. 

\begin{figure}
    \centering
    \includegraphics[width=0.9\linewidth]{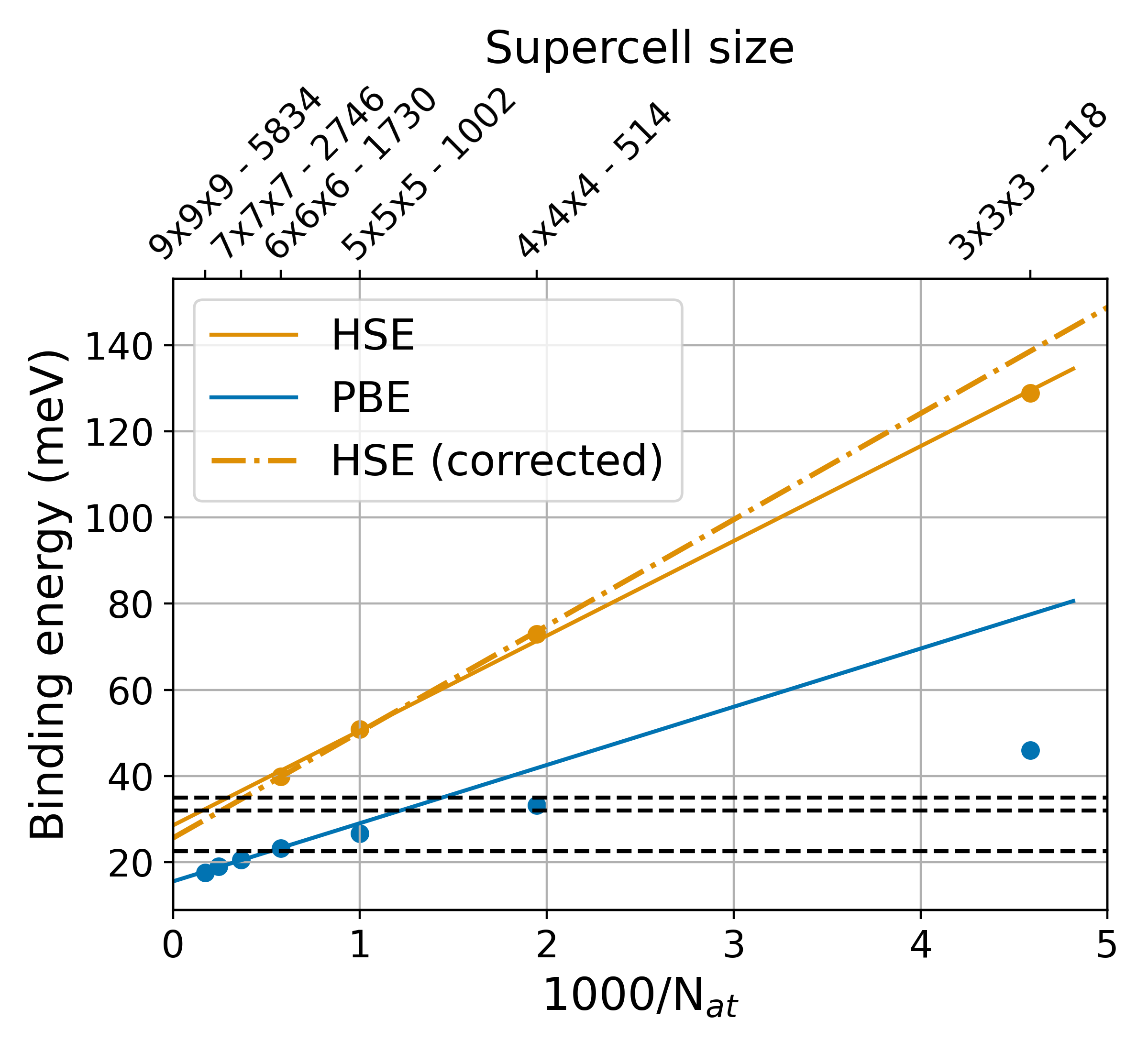}
    \caption{Calculated binding energy as a function of the cell size for PBE (blue) and HSE (orange) XC functionals. The corrected HSE slope is also shown (dash-dotted orange line). The dotted black lines are the reported experimental binding energies at 22.5, 32meV and 35meV\cite{Irion1985, Safonov1996, Bergeron2020}. The top axis indicates the supercell size and corresponding number of atoms. The 8x8x8 supercell tick was removed for clarity. }
    \label{fig:binding_energies}
\end{figure}

The calculated binding energy as a function of the cell size and for two exchange-correlation functionals, the semi-local functional from Perdew, Burke and Ernzerhof \cite{PBE} (PBE) and the hybrid from Heyd, Scuseria and Erzherhof \cite{HSE} (HSE), are shown in Figure \ref{fig:binding_energies}. For PBE and a supercell of size N$>$5 (N is the supercell size N$\times$N$\times$N), a clear linear trend is obtained when plotted against 1000/N$_{at}$, giving a binding energy of 14.62$\pm$1.37meV for N$\rightarrow \infty$ while a similar trend is obtained at smaller cell size for HSE, with an extrapolated binding energy of 28.51$\pm$5.53meV. The extrapolated binding energy with HSE is almost twice as large than the one obtained with PBE. As noted by Swift et al., this discrepancy is caused by the delocalization error of PBE, which decreases the electrostatic interaction between the hole and the charged defect \cite{Swift2020, Bryenton2023}. On the other hand, the high computational cost associated with HSE limits the maximal attainable cell size, thus the error from the fit. The 95\% confidence interval are represented by the filled region in Figure \ref{fig:binding_energies}. We are thus faced with a dilemma with PBE being more precise while HSE is more accurate. Swift et al. devised a methodology to reduce the fitting error of the HSE slope $A_{HSE}$ by using the PBE binding energies. The idea here is that the PBE fitted slope of the binding energy $A_{PBE}$ is correct for large supercells if it wasn't for the underestimation of the exchange term \cite{Swift2020}. If this error is corrected, then the HSE and PBE slopes, $A_{HSE}$ and $A_{PBE}$, should essentially be the same. More details about the methodology can be found in our Supplementary Material. Applying this approach to the T center, we obtain a corrected binding energy $E_{b,corr}$=25.59$\pm$0.69meV by extrapolating to infinite cell size. This value is very close to what is obtained by direct fitting to the HSE binding energy, supporting the validity of our scheme. For reference, these two methods are in very good agreement with the experimental binding energies which range between 22 to 35 meV \cite{Irion1985, Safonov1996, Bergeron2020}. This also illustrates the importance of such careful convergence study. A common size of supercell for silicon (or diamond) defects is 512 atoms. In our case, this supercell size would give a calculated binding energy around 72.90meV, which is about three times higher than the extrapolated exciton binding energy. 

\subsection{Stark shift}
The Stark shift, \textit{i.e.} the shift of the ZPL in the presence of an electric field, is caused by the coupling of the dipole moment change upon electronic transition. It is typically described by the expansion up to second order of the change in energy of the transition $\Delta E$ with respect to the electric field $\bm{F}$:
\begin{equation}
    \Delta E = \Delta \bm{\mu}  \bm{F} + \frac{1}{2} \bm{F} \Delta \bm{\alpha} \bm{F}
\end{equation}

For non-centrosymmetric geometries as is the case here, the linear response is expected to dominate the response.  Several approaches have been developed to calculate this including electric field-energy fitting \cite{Maze2011, Bathen2020, Alaerts2024}, modern theory of polarization \cite{Alaerts2024} and Wannier functions \cite{Lopez2024}. The electric field-based approach requires the construction of a slab model in which the defect is embedded. A slab is not the most adequate choice to model a bound exciton defect, our case here, because of the delocalized nature of the wavefunction which can lead to spurious interactions with the surface. More plainly, it is more difficult to contain the exciton in a slab geometry without a very large slab, necessitating a very large simulation box when vacuum is included. We therefore choose to work with the modern theory of polarization which can directly be applied to supercell calculations. In a previous work, we showed that this method is perfectly suitable to be applied on defects \cite{Alaerts2024}. It should be noted that within the modern theory of polarization formalism, dipole moments are only defined with respect to a centrosymmetric reference and modulo a quantum of polarization \cite{Spaldin2012}. Here, we take a difference between the ground state and excited state which alleviates the need for a centrosymmetric reference while the quantum of polarization, equal to $e \mathbf{R}$ where $\mathbf{R}$ is a lattice parameter, is very large compared to the calculate dipole moment change and can therefore be removed manually. 

\begin{figure}
    \centering
    \includegraphics[width=0.85\linewidth]{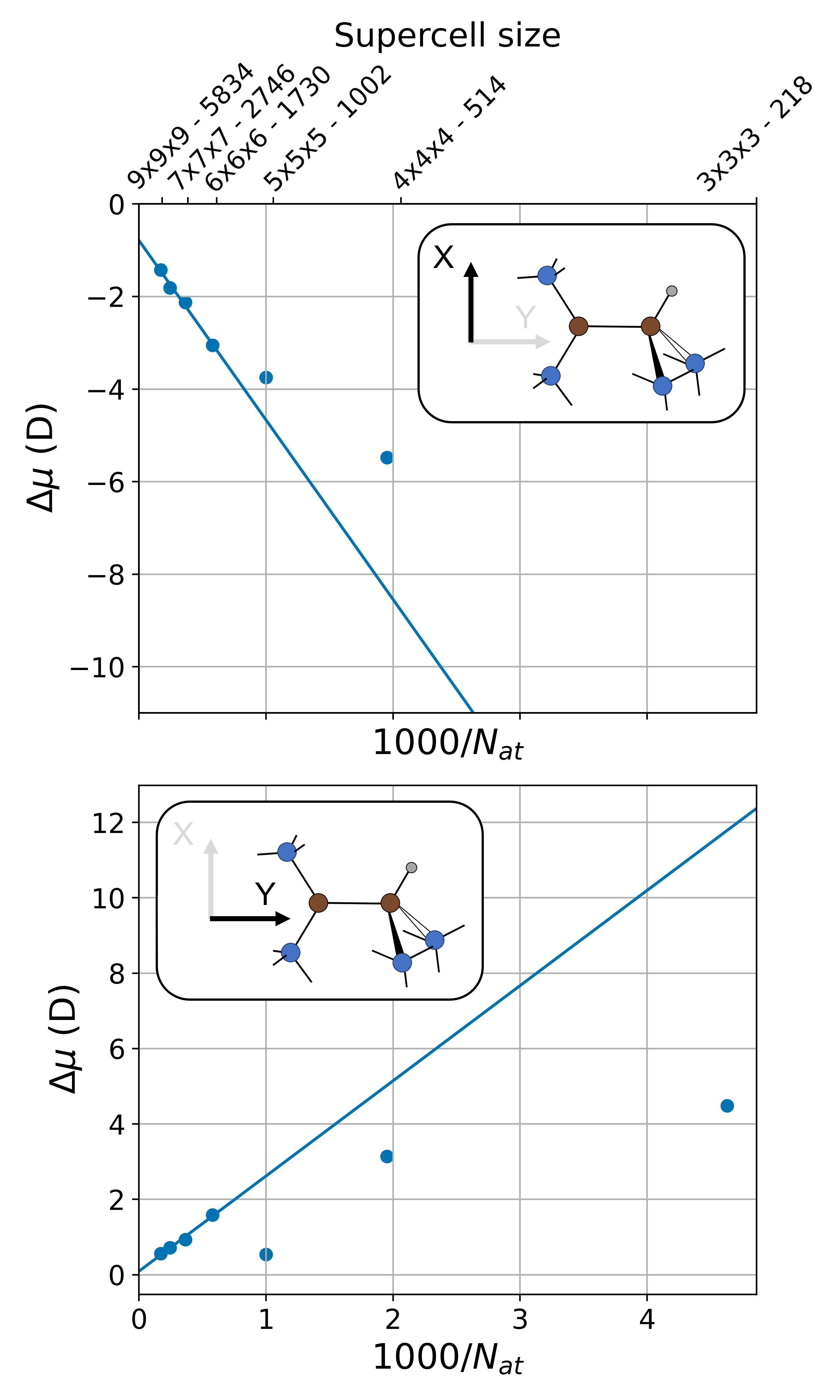}
    \caption{Dipole moment change as a function of the cell size calculated using PBE for the X component (top) and Y component (bottom). The top axis indicates the supercell size and corresponding number of atoms. The 8x8x8 supercell tick was removed for clarity.}
    \label{fig:stark_shift}
\end{figure}

Figure \ref{fig:stark_shift} shows the computed the dipole moment changes as a function of the supercell size using the PBE functional. We report the dipole moment in the defect basis, with the Y axis along the C-C axis, pointing towards the carbon bonded with H, and the X axis is in the C-C-H plane and is directed towards the H aside from a slight tilt. Due to the C$_{1h}$ symmetry of the defect, only $\mu_X$ and $\mu_Y$ have a non-zero dipole moment (i.e., $\mu_Z$=0). For large supercells (N$_{at} >$ 1000), a linear trend emerges and enable us to extract $\Delta \mu_{X, PBE}$=-0.79 and $\Delta \mu_{Y, PBE}$=0.09. Smaller supercells were excluded from the fit due to the large deviations with respect to the calculated slopes. For HSE, the small number of cell size leads to large uncertainties for both $\Delta \mu_{X}$ and $\Delta \mu_{Y}$, and larger supercells would be required to obtain a converged value (see Supplementary Material). We note that dipole moments calculated at the PBE and HSE levels are usually expected to be similar \cite{Zhang2017}. For instance, our own work on the NV center and using the same formalism gave $\Delta \mu_{PBE}$=2.68D and $\Delta \mu_{HSE}$=2.23D \cite{Alaerts2024}. Similar observations were made for the calculation of the dipole moment of molecules (see Supplementary Material) \cite{Hickey2014, Hait2018}. We therefore expect that semi-local functionals such as PBE are capturing most of the physics required to get accurate dipole moment changes, as long as the wavefunction is entirely contained within the simulation cell. 

We can estimate the extend of the bound exciton wavefunction by computing the exciton Bohr radius $a^*$ using the projected charge density of the hole state. If the defect results in a spherically symmetric potential as it is the case for simple acceptors such as boron or aluminum, we could fit the spherically averaged charge density and extract $a^*$ from the decaying charge density \cite{Smith2017}. However, it is clear that the crystal field potential of the T center is far from spherical and instead, we evaluate the exciton Bohr radius for three different crystallographic planes, [110], [$\Bar{1}$10] and [001] (see Supplementary Material). [$\Bar{1}$10] is the plane in which the T-center lies. Using the largest supercell ($N_{at} = 5834$) and PBE, we obtain decaying lengths of 34.99\r{A}, 23.83\r{A} and 21.77\r{A}. The longest Bohr radius, 34.99\r{A} is found in the [$\Bar{1}$10] plane, along the [1,1,1] crystallographic axis. Thus, we can expect that cell size with dimensions smaller than $\approx$ 40\r{A}, which is the case for all our HSE calculations, are not entirely accurate. On the other hand, for PBE we have a set of calculations respecting this criterion. In fact, it is only around these cell dimensions that our expected linear trend emerges. 

The knowledge of the Bohr radius of the exciton also allow us to estimate the quadratic response assuming it is proportional to the volume of the wavefunction and that the contribution of the ground state is negligible \cite{Tamarat2006, Clear2024}. The volume of an ellipsoid is given by the product of the three decaying lengths, giving a wavefunction volume of 7.6 $\times$10$^4$\r{A}$^3$. The corresponding polarizability change $\Delta \alpha$ = 0.0127Hz.m$^2$/V$^2$ is obtained by multiplying the wavefunction volume by 4$\pi$$\varepsilon_0$ \cite{Tamarat2006}. This is significantly larger than the value we recently calculated for the NV center for which we obtained $\Delta \alpha$ $\approx$ 5$\times$10$^{-6}$Hz.m$^2$/V$^2$ or $\approx$ 20 \r{A}$^3$ in polarizability volume \cite{Alaerts2024}.

\begin{table}[h]
\centering
\renewcommand{\arraystretch}{1.3} % Adjust row height
\setlength{\tabcolsep}{5pt} % Adjust column spacing
\begin{tabular}{ccccc}
\hline
\hline
                 & $\Delta \mu_X$ & $\Delta \mu_Y$ & $\alpha_{XX}$ & $\alpha_{YY}$ \\
\hline
\cite{Clear2024}      & -0.71             & 1.49               & 0.123                        & 0.106                                 \\
This work             & -0.79             & 0.09               & \multicolumn{2}{c}{0.0127}             \\
\hline
\hline
\end{tabular}

\caption{Comparison between experimental and theoretical dipole moment and polarizability changes of the ZPL transition \cite{Clear2024}. The dipole moments are given in D and the polarizability in Hz.m$^2$/V$^2$}
\label{tab:comparison_stark_shift}
\end{table}

\subsection{Effect of the delocalized hole on the Stark shift}
\begin{figure}
    \centering
    \includegraphics[width=0.95\linewidth]{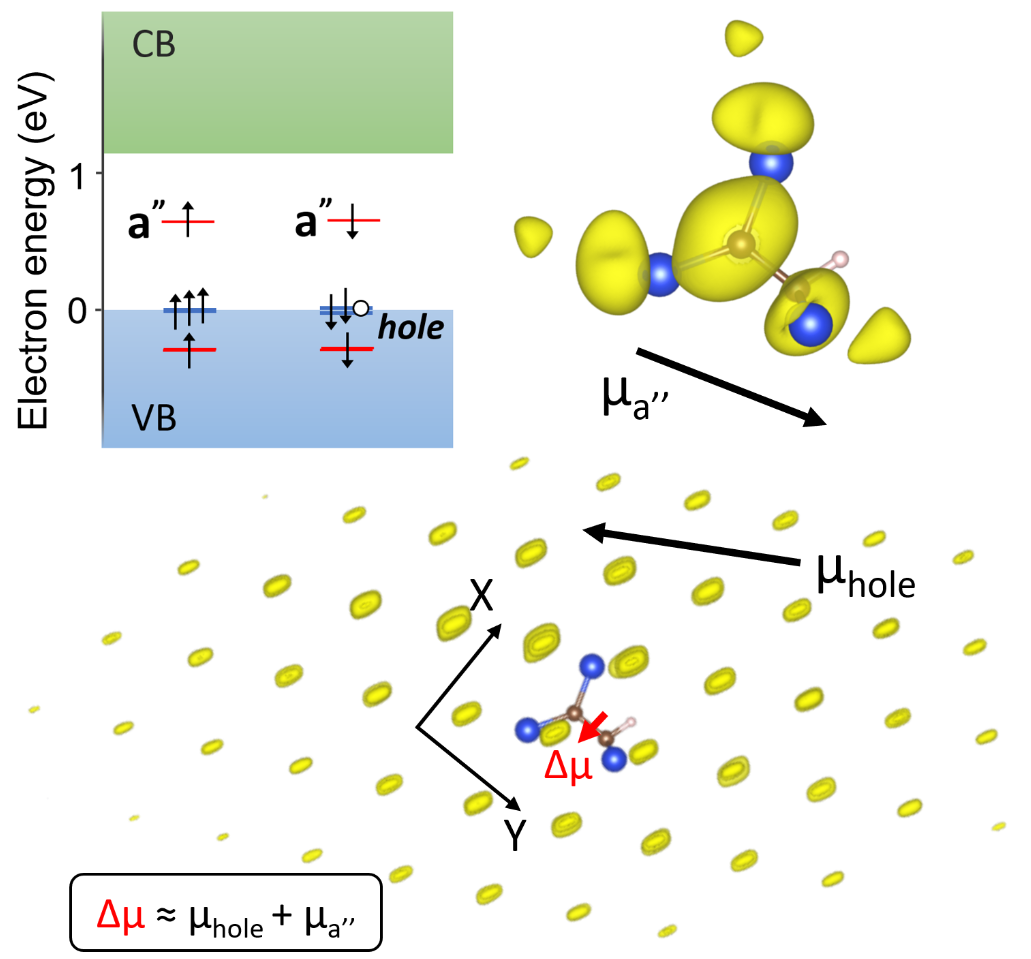}
    \caption{Exciton contribution to the dipole moment change as a sum of two process, the accepting process (the occupation of the a'' state) and the formation of a hole in the valence band. The top right figure shows the charge density contour plot of the a'' level along the [-110] crystallographic direction. The bottom part shows the same for the hole exciton. The defect coordinate basis is indicated as well.}
    \label{fig:exciton_contribution}
\end{figure}

Finally, we assessed the individual contributions of the exciton to the calculated dipole moment change. The approach used to calculate the dipole moment of the excited state requires a decomposition band-by-band \cite{Alaerts2024}, thus giving direct access to the contribution of the hole and a'' levels (see top left, Figure \ref{fig:exciton_contribution}). Such decomposition was previously used in the study of ferroelectric materials \cite{Ghosez1995}. Within a hydrogen model and excluding spin-orbit coupling, the dipole moment of the hole is zero because of the spherical symmetry of the wavefunction. Taking into account the symmetry of the impurity or the effect of spin-orbit coupling, the dipole moment should be non-zero but we would naively expect the effect of these perturbations to be rather small \cite{Smit2004}. Surprisingly, we find here that the hole contribution $\mu_{hole}$ has a rather large dipole moment: -2.03D and -3.47D for the X and Y components, respectively. Similarly, the contribution of the accepting process $\mu_{a''}$ is obtained from the dipole moment of the a'' state. We find $\mu_{a''}$ to be +1.12D along X and +3.50D along Y. Figure \ref{fig:exciton_contribution} summarize our analysis. This suggests that the small dipole moment change associated with the ZPL transition emerges as the cancellation of two large quantities, $\mu_{hole}$ and $\mu_{a''}$. The sum of the two process, -0.91D along X and 0.03D along Y, is shown by the red arrow in Figure \ref{fig:exciton_contribution} and is very close to what we obtained for $\Delta \mu$. Alternatively, we can obtain the exciton contribution by assuming the excitation takes place in two stages: first, the T center becomes negatively charged and then the bound exciton is formed when the hole binds to the negative defect. In this case, we obtained $\mu_{T-}$ = +2.32D and +5.56D and $\mu_{hole}$=$\Delta \mu - \mu_{T-}$ = -3.08 and -5.47 for X and Y, respectively. This corresponds to a exciton contribution of 0.76D along X and 0.09D along Y. The difference between the two schemes is due to electronic relaxation effects when calculating the dipole moment of the negative T center but leads to essentially the same conclusion. 

The dipole moment change associated with the formation of the exciton can also be understood from the charge density redistribution associated with the charging process. When an electron is captured by the T center, it will mostly be localized on the dehydrogenated carbon as shown in the top right part of Figure \ref{fig:exciton_contribution}. Assuming the electronic density is initially located around the center of mass of the T center, which is supported by the very small dipole moment of the T center ground state, the localization of a charge of -$e$ onto the carbon corresponds to a transfer of charge of -2.45D along X and +5.00D along Y. The real space visualization of the exciton wavefunction helps understanding its influence on the dipole moment change (see Supplementary Material). The close to perfect compensation along the Y direction is due to the orientation of the defect with respect to the crystallographic axis. Exciton wavefunctions are typically expressed as the product of a decaying envelope and of the VBM. Here, the crystal field splitting introduced by the T center breaks the degeneracy of the VBM and constrain the exciton wavefunction mostly along the [001] crystallographic direction, which is almost perfectly commensurate with the defect Y axis. We note that this is in contrast with group III acceptors such as boron, for which the valence bands stay degenerate (excluding the effect of spin-orbit coupling) and the resulting exciton wavefunction is quasi-spherical \cite{Wang2009} (see Supplementary Material). In this case, we would therefore expect that the dipole moment change associated to the hole wavefunction is relatively small. For comparison, the linear Stark shift of B in Si has been estimated between 0.16D and 0.87D \cite{Kopf1992, Calvet2007}. 

\subsection{Comparison with experiments}
The linear and quadratic coefficient of the Stark shift for the T center was recently measured by Clear et al. \cite{Clear2024}, enabling a comparison between theory and experiment (see Table \ref{tab:comparison_stark_shift}). For the dipole moment change, we have a very good agreement for the component along X but a large discrepancy for Y. Our calculations predict a linear response close to zero, clearly below the experimental value of $\Delta \mu_{Y}$=+1.49D. This is surprising considering the good agreement between the dipole moment of molecules calculated with DFT and their experimental values. For PBE, the largest absolute deviation found in Ref. \cite{Hickey2014} is 0.548D for cytosine but its relative deviation is only -8\%, far lower than what we observe here. The accuracy of DFT is thus unlikely to explain this discrepancy and instead, we hypothesize it is related to local field effects. In an experimental measurement, the applied voltage has to be converted to a local field which itself depends on many variables such as the presence of impurities in the vicinity of the defect. This is particularly important to consider since the measurements were performed on T center ensembles. As we have seen above, the delocalized nature of the exciton and its corresponding large volume leads to large polarizability change, meaning that induced dipole moments arising from the interaction with other nearby impurities could be quite large. A quick estimation of the induced dipole by a charged impurity shows that this effect can be on the order of 1D for an impurity located at 50nm (see Supplementary Material). The impact of nearby impurities on the measured Stark shift has been observed in the case of P donors in silicon \cite{Guichar1972}. This suggests that bound-exciton defects could be intrinsically sensitive to local field effects as a result of their large spatial extent. It is important to note that these effects are not as important in defects with very localized levels such as the NV center in diamond. For reference, our estimate for the quadratic response is one order of magnitude below the experimental value but clearly supports the model of a large polarizable defect. It should be noted that in our calculations, we effectively remove the effect of these spurious interactions by performing a convergence study. Finally, our small calculated Stark shifts are consistent with a recent report of weak optical tuning for T center integrated in a p-i-n junction \cite{Day2025}. It would be interesting to compare our results to measurements performed on individual T center.  

Our work also raise an important question about the stability of the exciton. If the applied electric field creates a potential drop larger than the binding energy over across the Bohr radius of the exciton, then the exciton will dissociate and quench the TX lines from the PLE spectrum. We calculate this critical field using $F_{crit}$=$\frac{E_b}{e a^*}$. Using our calculated numbers of $E_B$=25.59meV and $a^*$=35.64\r{A} (for the longest axis), we obtain $F_{crit}\approx$70kV/cm. This value is about two orders of magnitude above the larger the fields applied in Ref. \cite{Clear2024}. Further, it seems that the heating caused by leakage currents at fields below the critical field can already be enough to thermally dissociate the exciton \cite{Day2025}.

\section{Conclusion}
We have calculated the Stark shift of the T center in silicon using first-principles calculations with a focus on the linear response. The technical challenge associated with the formation of a delocalized defect-bound hole-exciton during the ZPL transition is tackled by performing a convergence study over the supercell size. We obtain a computed exciton binding energy of approximately 25 meV, in good agreement with experimental values. For the calculation of the Stark shift, changes in dipole moments were calculated within the modern theory of polarization formalism. A careful convergence with the unit cell size leads to $\Delta \mu_{X}$=-0.79D and $\Delta \mu_{Y}$=+0.09D using PBE. Our calculation also reveals that this small dipole change emerges as the cancellation of two large quantities associated with the formation of a hole and the occupation of the defect a'' level. This value is smaller than the NV center in diamond using the same level of theory (2.68D). This value is also smaller than measured in recent ensemble measurements. We hypothesize that the source of this discrepancy could be associated with local field effects arising from nearby charged impurities in experiments. Our results indicate that the sensitivity of the T center ZPL emission wavelength to electric fields is reasonable compared to other quantum defects such as the NV center in diamond or the G center in silicon \cite{Day2024} but could strongly depends on the electrical environment of the defect.

\section{Methods}
We used the Vienna ab-initio simulation package (VASP) to conduct all our calculations. VASP is a a density functional theory (DFT) plane-wave code \cite{DFT1, VASP1, VASP2} based on the projected augmented wave method formalism \cite{PAW1}. The two exchange-correlation functionals used in this work are the formulation of Perdew-Burke-Ernzerhof of the generalized gradient approximation (GGA-PBE) \cite{PBE} and the hybrid Heyd-Scuseria-Ernzherof with 25\% of exact exchange (HSE) \cite{HSE}. Due to the large cell size, the reciprocal space was sampled using a $\Gamma$-only grid. The energy cutoff for the plane-wave basis set was set to 400eV and 520eV for the HSE and PBE calculations, respectively. Unless specified otherwise, the self-consistent field loop was stopped once the difference in energy between two consecutive steps was less than 10$^6$ eV. When an ionic relaxation was performed, the convergence was stopped once the forces on the atoms were less than 0.01eV/\r{A}. 

The T center was embedded at a center of a NxNxN supercell. For PBE, we performed calculations on size ranging from N=3 (218 atoms per cell) to N=9 (5834 atoms per cell). All the atomic degrees of freedom were relaxed for cell size up to N=6 (included). For larger cell size, the geometry of the cell size N=6 was used. For HSE, the size ranged from N=3 to N=6 (5834 atoms per cell) and relaxation were done up to N=5. The electronic occupation of the excited was constrained using the  $\Delta$-SCF method. The calculation of the dipole moment using the modern theory of polarization usually requires a centrosymmetric reference to avoid the inclusion of a quantum of polarization. Here, we did not use such a reference because the dipole moments are much smaller than the quantum of polarization which is $\approx$ 79D. Due to an issue with the VASP code, the calculation of dipole moment in the excited state requires special care and we use the methodology of Alaerts et al. \cite{Alaerts2024}. All our calculations were prepared and analyzed using Pymatgen \cite{Ong2013}. 

\begin{acknowledgments}
We thank Alp Sipahigil, Chloe Clear, Aaron Day for fruitful discussions. This work was supported by the U.S. Department of Energy, Office of Science, Basic Energy Sciences 
in Quantum Information Science under Award Number DE-SC0022289.
This research used resources of the National Energy Research Scientific Computing Center, 
a DOE Office of Science User Facility supported by the Office of Science of the U.S.\ Department of Energy 
under Contract No.\ DE-AC02-05CH11231 using NERSC award BES-ERCAP0020966. 
\end{acknowledgments}

\bibliography{biblio}% Produces the bibliography via BibTeX.

\clearpage

\title{Supplementary Material}
\author{}
\date{}
\newcommand{\supplementarytitle}[2][]{
  \clearpage
  \thispagestyle{plain}
  \begin{center}
    {\LARGE \bfseries Supplementary Material\par}
    \vspace{0.5em}
    {\large #2\par}
    \vspace{1em}
    {\small #1\par}
  \end{center}
}
\supplementarytitle

\setcounter{section}{0}
\renewcommand{\thesection}{S\arabic{section}} % Optional: S1, S2, ... format

\section{Correcting the HSE binding energy with PBE}
The binding energies calculated at the PBE level grossly underestimate the experimental value because of the delocalization error associated with the handling of the exchange potential. This error is corrected by the HSE hybrid functional but its higher computational cost forbids its usage to large cell sizes. In consequence, if the error arising from the PBE exchange is corrected, then the fitted slope of the binding energy calculated at the PBE and HSE, $A_{PBE}$ and $A_{HSE}$, level should essentially by the same. We evaluate the PBE error on the exchange energy by measuring the behavior of the exchange splitting, $\delta_{ex}$ and defined as the energy difference between the spin-up and spin-down eigenvalues of the acceptor state, with respect to the cell size. Naturally, the corresponding slopes at the PBE and HSE level, $\delta^{fit}_{PBE}$ and $\delta^{fit}_{HSE}$, are different. The correct binding energy slope $A_{corr}$ is obtained by subtracting half of the exchange splitting $\delta_{ex, PBE}$ to the PBE binding energy slope $A_{PBE}$, thus removing the dependency of $A_{PBE}$ on the exchange, and by adding the exchange splitting obtained using HSE $\delta^{fit}_{HSE}$: $A_{corr} = A_{PBE} - \frac{1}{2}\delta^{fit}_{PBE} + \frac{1}{2}\delta^{fit}_{HSE}$. In other words, we have replaced the (fallacious) PBE exchange with the (correct) HSE exchange.

\section{Convergence study on the dipole moment change with HSE}
\begin{figure}[!h]
    \centering
    \includegraphics[width=0.95\linewidth]{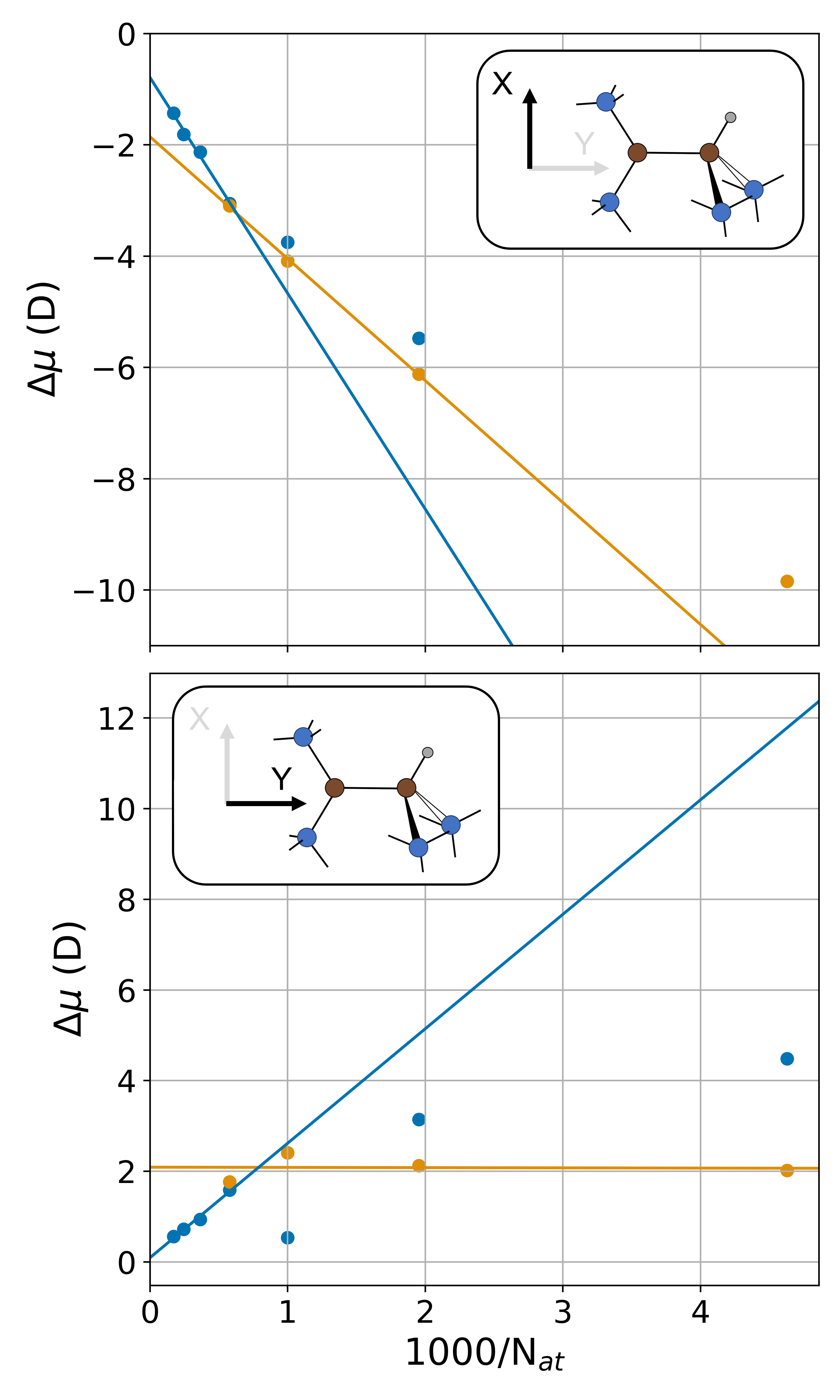}
    \caption{Dipole moment change as a function of the cell size calculated using PBE (blue) and HSE (orange) for the X component (left) and Y component (right).}
    \label{fig:SM_stark_shift}
\end{figure}

We calculated the dipole moment changes of the ZPL transition of the T center as a function of the cell size using both PBE and HSE (see Figure \ref{fig:SM_stark_shift}). The dipole moment changes for an infinite cell size are obtained by extrapolation. For PBE, a linear trend is only obtained for cell size larger than $N_{at} >$ 1000 atoms, allowing us to extrapolate $\Delta \mu_{X, PBE}$=-0.79 and $\Delta \mu_{Y, PBE}$=0.09. The situation is a more complicated for HSE due to the limitations in terms of cell size. For $\mu_X$ and N$>$3, we recover an apparent linear trend and the extrapolated slope gives $\Delta \mu_{X, HSE}$=-1.85. On the other hand, the convergence is much less obvious for $\Delta \mu_{Y, HSE}$. Including all the calculated dipole moments into the fit leads to $\Delta \mu_{Y, HSE}$=2.09D while excluding the two smaller cells gives $\Delta \mu_{Y, HSE}$=0.89D (not shown). Clearly, it is difficult to obtain a definite value because of the limitation in terms of cell size.

\section{Comparison between PBE and HSE}
Here, it is difficult to conciliate the PBE and HSE dipole moment changes even when the large uncertainty is taken into account. At first glance, the situation seems similar to what we have obtained for binding energies and one might be tempted to choose the HSE results as the correct ones. However, contrary to binding energies, dipole moments calculated at the PBE and HSE levels are usually expected to be similar \cite{Hickey2014, Hait2018, Zhang2017, Alaerts2024}. This is illustrated in Table \ref{tab:PBE_vs_HSE_molecules} for a set of molecules found in ref. \cite{Hickey2014}. For defect calculations, the only comparison found in the literature is our own work on the NV center. For the modern theory of polarization calculation, the dipole moment change calculated using PBE is $\Delta \mu_{PBE}$=2.68D while $\Delta \mu_{HSE}$=2.23D. 

% Please add the following required packages to your document preamble:
% \usepackage{booktabs}
% \usepackage{graphicx}
\begin{table}[]
\centering
\caption{Comparison between the calculated dipole moment at the PBE/aug-cc-pVTZ and PBE0/cc-pVTZ level with the for a set of molecules taken from Hickey et al. \cite{Hickey2014}}
\label{tab:PBE_vs_HSE_molecules}
\begin{tabular}{@{}ccccc@{}}
\toprule
name              & formula & exp   & PBE   & PBE0  \\ \midrule
carbon monoxide   & \ce{CO}  & 0.11  & 0.193 & 0.145  \\
nitrogen monoxide & \ce{NO}  & 0.159 & 0.222 & 0.121 \\
water             & \ce{H2O}  & 1.855 & 1.804 & 1.924 \\
methanol          & \ce{CH4O}  & 1.7   & 1.587 & 1.589 \\
acetone           & \ce{C3H6O}  & 2.88  & 3.001 & 2.929 \\
acetaldehyde      & \ce{C2H4O}  & 2.75  & 2.785 & 2.740 \\
acetic acid       & \ce{C2H4O2}  & 1.7   & 1.736 & 1.719 \\
dimethylamine     & \ce{C2H7N}  & 1.01  & 0.922 & 0.905 \\
pyrrole           & \ce{C4H5N}  & 1.74  & 1.856 & 1.937  \\
chlorobenzene     & \ce{C6H5Cl}  & 1.69  & 1.682 & 1.731 \\ \midrule
RMSD              &         &       &  0.137     &   0.137  \\
\bottomrule
\end{tabular}
\end{table}

\begin{figure}[!h]
    \centering
    \includegraphics[width=0.95\linewidth]{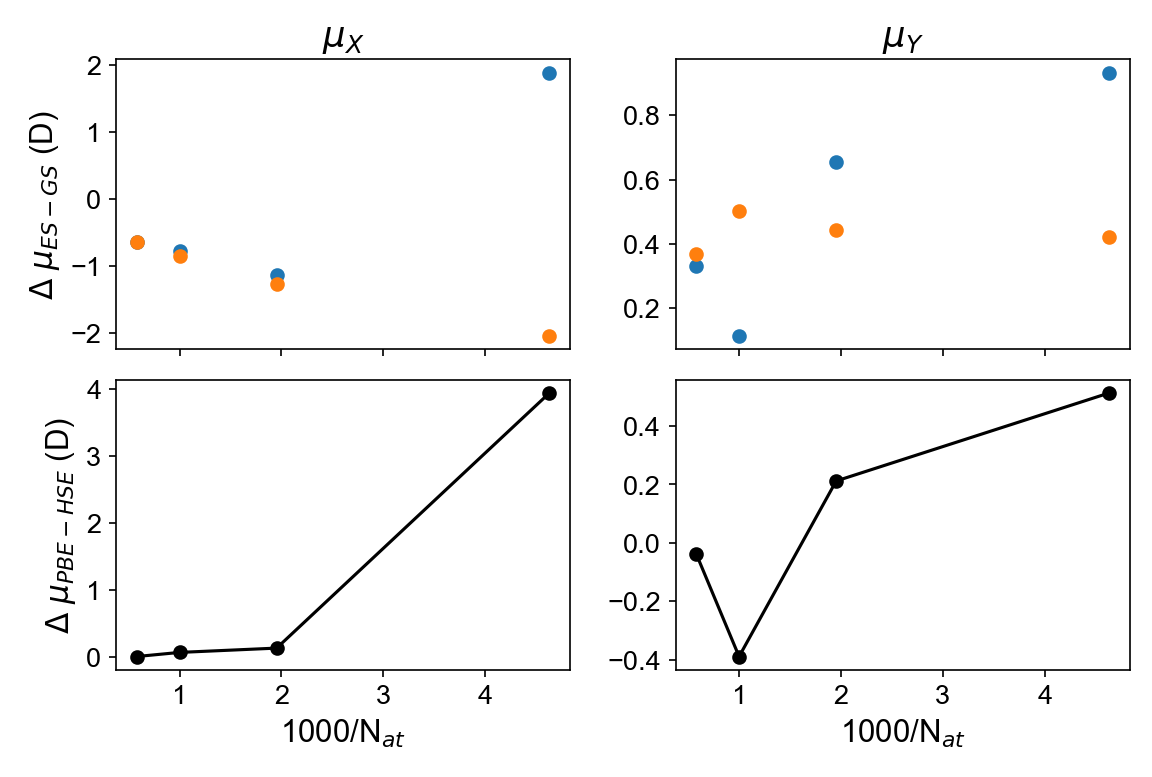}
    \caption{Dipole moment changes as a function of the cell size calculated using PBE (blue) and HSE (orange) for the X component (top left) and Y component (top right). Difference between the dipole moment change calculated at the PBE and HSE levels for the X component (bottom left) and Y componennt (bottom right).}
    \label{fig:SM_delta_PBE_HSE}
\end{figure}

Another argument that leads us to believe the PBE dipole moment is correct is that the discrepancy between the PBE and HSE dipole moment changes tends to decrease for larger supercells and is almost zero for the largest supercell calculated at the HSE level (see Fig \ref{fig:SM_delta_PBE_HSE}).

\section{Exciton radius and contribution to the dipole moment}
\begin{figure}[!h]
    \centering
    \includegraphics[width=1\linewidth]{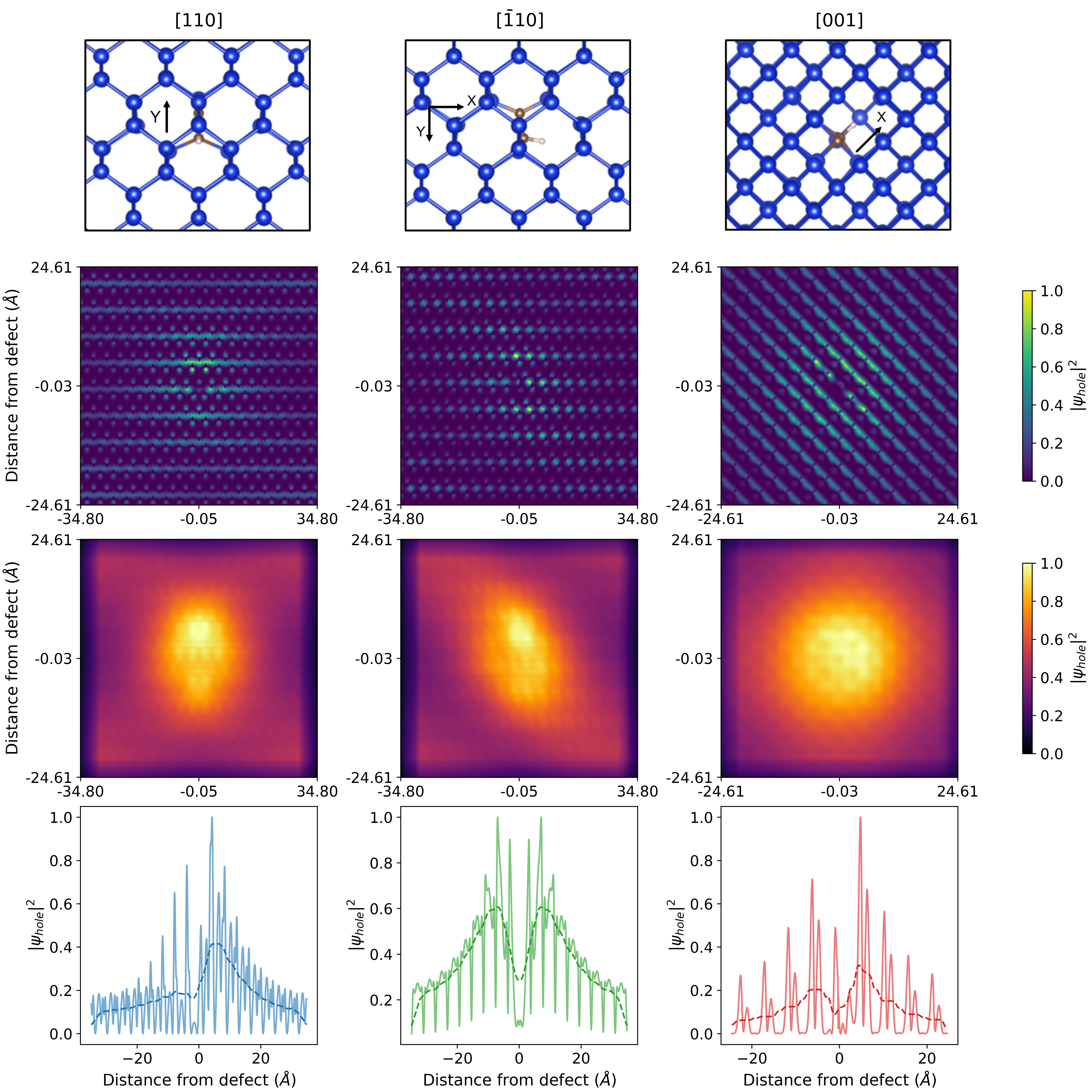}
    \caption{On the top row, we show the orientation of the defect with both the crystallographic axis and the defect coordinate system. The second row shows hole charge density, $|\psi_{hole}|^2$ slices taken along different crystallographic plane while the third row shows the convoluted charge density. The effect of a convolution on the charge density is to remove the periodic part. In other words, we isolate the decaying exponential envelope function. The last row shows string of charge density. The solid line represents the actual charge density while the dashed line is the convoluted charge density.}
    \label{fig:exciton_wf}
\end{figure}

\begin{figure}
    \centering
    \includegraphics[width=1\linewidth]{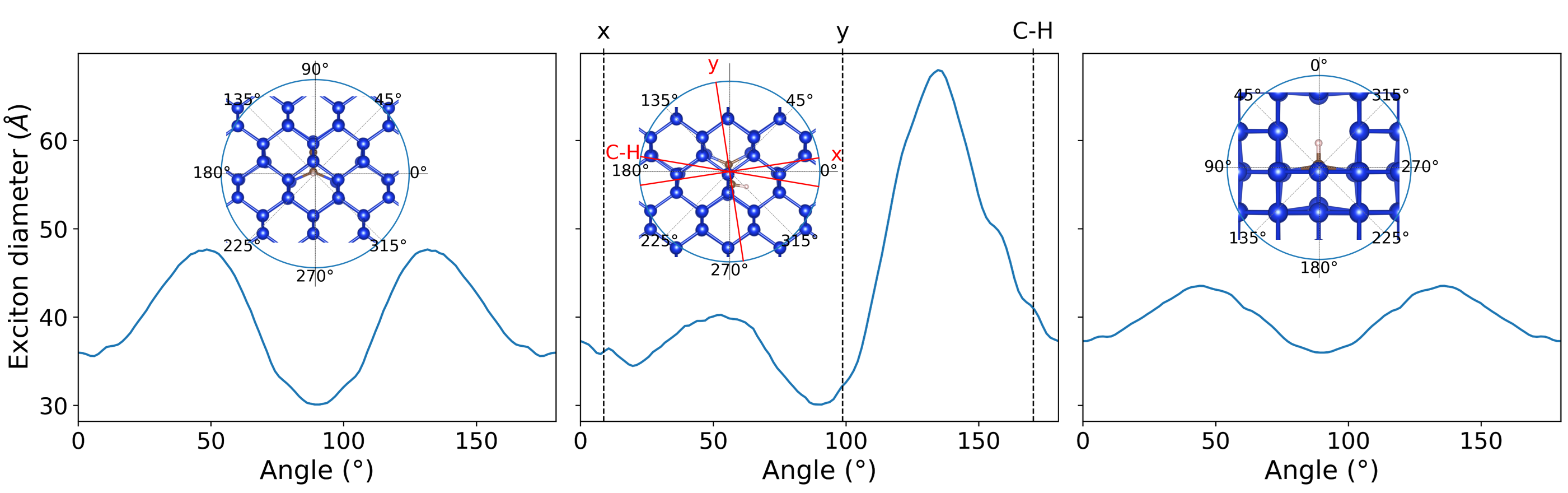}
    \caption{Estimation of the exciton diameter around three different crystallographic orientation, [110] (left), [$\bar{1}$10] (center) and [001] (right).}
    \label{fig:exciton_diameter}
\end{figure}

The problem of finding the Bohr radius of the exciton state of the T center is more complicated than in simple acceptors such as boron because of the low defect symmetry. For boron, one can assume that beyond a cut-off distance, the negatively charged defect leads to a spherically symmetric potential and thus, a simple exponential $F(r) = A \, \mathrm{exp}(\frac{-r}{a_0})$ can be used to fit the decaying envelope wavefunction \cite{Smith2017}. In practice, this is done by taking the spherical average of the wavefunction. 
Here, we choose an alternative approach. First, we apply a convolution to the charge density corresponding to the hole state to remove the periodic part and isolate the decaying exponential envelope function. This operation is shown on Figure \ref{fig:exciton_wf} for three specific crystallographic planes, [110], [$\bar1$10] and [001]. For each plane, one-dimensional strings of charge density are then selected around a circle whose center is chosen as the hydrogen-bonded carbon. For each of these charge density strings, we fit a decaying exponential $F(r) = A \, \mathrm{exp}(\frac{-r}{a_0})$ to extract the Bohr radius $a_0$.
The exciton diameter for a specific plane and angle is then obtained by summing the Bohr radius in each direction. The result of this analysis is shown on Figure \ref{fig:exciton_diameter}. The longest calculated Bohr diameter is 69.97\r{A} and is found in the [$\bar{1}$10] plane, which corresponds to the plane defined by the C-C-H atoms, along the [111]  crystallographic axis. Thus, in order to fully encompass the exciton, the cell dimension must have a length of at least 69.97\r{A} along the [111] axis, or a lattice constant of 40\r{A}.  For the [110] and [001] plane, the longest diameters are 47.66 and 43.54 \r{A}, respectively.

In the main text, we have seen that the exciton contribution to the dipole moment is $\mu_{X}$ = +2.025D and $\mu_{Y}$ = -3.470D. This is consistent with the real space visualization of the exciton shown in Figure \ref{fig:exciton_wf}. This is particularly clear from the [$\bar{1}$10] projection which shows both the spatial distribution of the hole charge density along the X and Y directions. 

\section{Comparison with B in Si}
\begin{figure}
    \centering
    \includegraphics[width=0.95\linewidth]{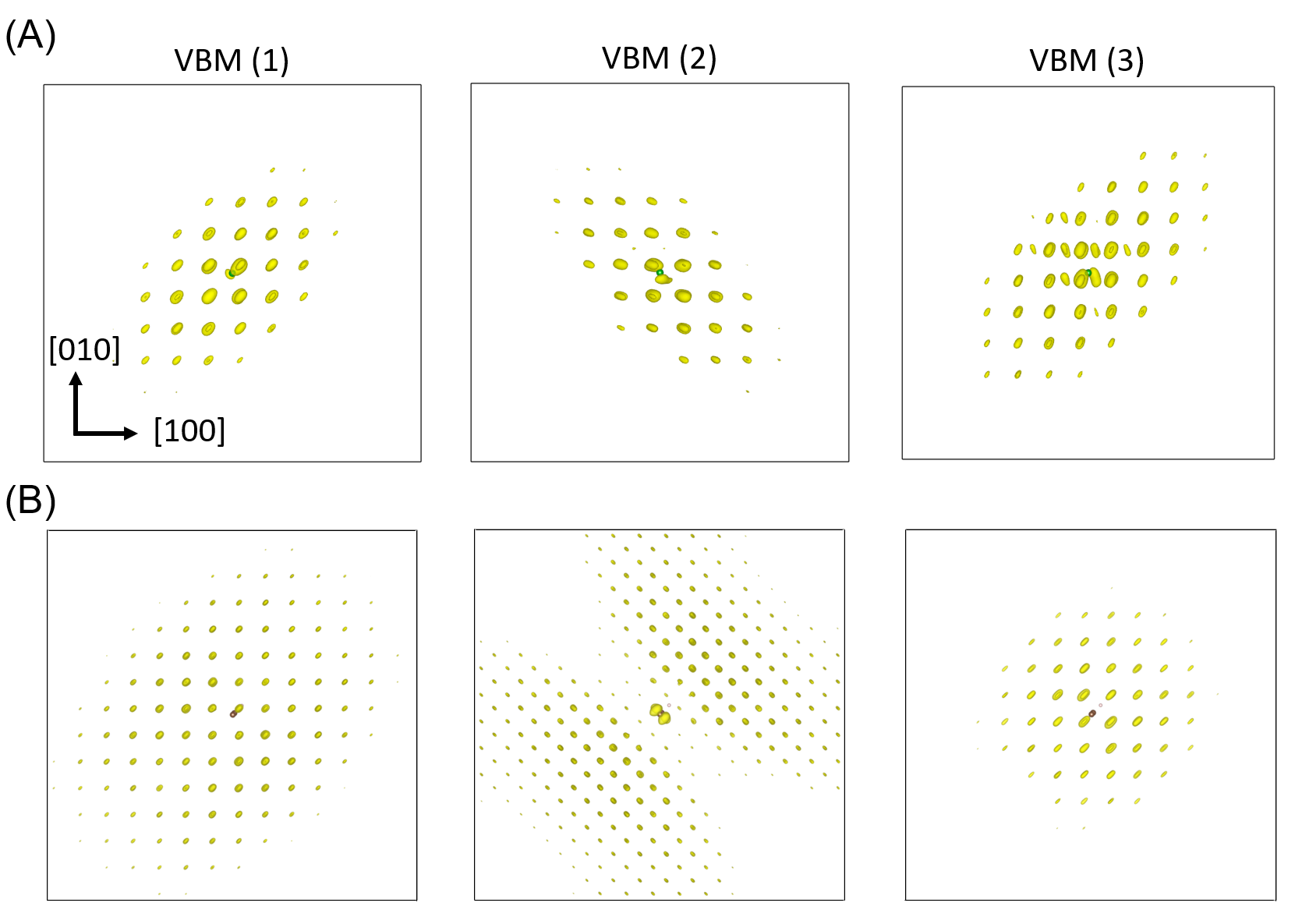}
    \caption{(A) Charge density isosurface of the three valence band of Si projected along the [001] direction for the case of a B dopant. The B dopant is shown in green, in the center of cell. Each wavefunction has a cigar shaped. (B) Same but for the T center. }
    \label{fig:B_vs_T}
\end{figure}

The similarity between the pseudo-acceptor state of the T center and group III acceptor such as B motivate us to directly compare their exciton wavefunctions. As discussed in the main text, when B is implemented into silicon, it introduces a shallow acceptor level which is easily occupy by a delocalized electron from the valence band. Because B occupies a Si site, it possess the Td point group and this means that, if spin-orbit is excluded, the valence band maximum stays triply degenerated. The wavefunction of the hole is thus the product of a hydrogen like radial function and of the linear combination of the three valence band maximum. On the other hand, the low symmetry of the T center, C$_{1h}$, will introduce a crystal field splitting. The triply degenerate VBM (again, excluding spin-orbit coupling) is now split into three levels. The energy split is on the order of 3-15meV while the experimental separation of the TX$_0$ and TX$_1$ is 1.8meV. Within this model, the hole exciton wavefunction of the T center comes only from the VBM which has a specific crystallographic orientation (see Figure \ref{fig:B_vs_T}) and is commensurate with the C-C axis.

\section{Dipole moment induced by impurities}
In this section, we show using a very simple model that the dipole moment induced by impurity can be quite large as a result of the important polarizability change of the T center. The induced dipole moment is directly given by the product of the polarizability with the electric field:

\begin{equation}
    \Delta \bm{\mu}_{induced} = \alpha \cdot \bm{F}
\end{equation}

Estimating the actual electric field felt by a T center is very difficult but we can assume the simplest case in which the impurities are perfectly distributed in the sample. In this case, the average distance between two impurities is given by $\sqrt[3]{\rho}$ where $\rho$ is the impurity concentration. As shown in Table \ref{tab:my-table}, the induced dipole moment can quickly become large, on the same order of magnitude than the dipole moment change for the ZPL transition. 

\begin{table}[]
\centering
\caption{Estimation of the induced dipole moment for different charged defect concentration. We used $\alpha$=0.123Hz.m/V \cite{Clear2024}.}
\label{tab:my-table}
\begin{tabular}{ccccc}
\toprule
$\rho$ (cm$^3$)    & d (nm)         & E (kV/m)    & $\Delta \mu$ (Hz.m/V)  &  $\Delta \mu$ (D)   \\ 
\midrule
1E+14              & 215.4435       & 2.65   & 325.85            &  0.06   \\
1E+15              & 100            & 12.30   & 1512.47            & 0.30    \\
1E+16              & 46.41589       & 57.08   & 7020.25            & 1.39    \\ \bottomrule
\end{tabular}%
\end{table}

\end{document}